\documentclass[12pt]{article}
\begin{document}
\title{$\kappa$-Deformed Kinematics and Addition Law
for Deformed Velocities\thanks{Supported by KBN grant
5PO3B05620}
\thanks{To appear in special issue of Acta Physica
Polonica B, dedicated to 60-th birthday of Stefan Pokorski}}
\author{Jerzy Lukierski\\
Institute of Theoretical Physics, University of Wroc\l aw \\ pl. M. Borna 9,
50-205 Wroc\l aw, Poland
\\ \\
Anatol Nowicki \\ Institute of Physics, University of Zielona
G\'{o}ra\\ pl. S\l owia\'{n}ski 6, 65-069 Zielona G\'{o}ra,
Poland}

\maketitle
\begin{abstract}
In $\kappa$-deformed relativistic framework we consider three
different definitions of $\kappa$-deformed  velocities  and
introduce corresponding addition laws. We show that one of the
velocities has classical relativistic addition law. The relation
of velocity formulae with the coproduct  for fourmomenta and
noncommutative space-time structure is exhibited.
\end{abstract}

\section{Introduction}
Recently due to increasing interest in deformed relativistic space-time
framework (see. e.g. [1-6]) it is important to understand the deformation of
Einsteinian relativistic kinematics. In particular the problems occur if the
classical Poincar\'{e} symmetries are replaced by quantum ones, with
modification of classical Abelian addition law for the momenta. Here we shall
consider as distinguished example the so--called $\kappa$-deformed quantum
Poincar\'{e} symmetries (see e.g. [7-11]), which recently were also used as
possible framework for describing the quantum gravity effects (see. e.g.
[12-16]).

The $\kappa$-deformed relativistic Hopf algebra framework in
bicrossproduct basis is characterized by classical Lorentz algebra
of $O(3,1)$ generators $M_{\mu\nu}=(M_i, N_i)$, commuting
fourmomenta $P_\mu=(P_i, P_0 = \frac{E}{c})$ and $\kappa$-deformed
commutation relations of threemomenta $P_i$ with boost generators
$N_i$:
\begin{equation}\label{1}
  [N_i, P_j] = \frac{i}{2} \delta_{ij}
  \left[ \kappa c (1 - e ^{ -\frac{2E}{\kappa c^2} })+ \frac{1}{\kappa
  c}\vec{P}^{\ 2}\right] - \frac{i}{\kappa c}P_i P _j
\end{equation}
All remaining Poincar\'{e} algebra relations remain classical. The mass square
Casimir $M^2$ for $\kappa$-deformed Poincar\'{e} algebra in bicrossproduct
basis  looks as follows [3,8,9]
\begin{equation}\label{2}
  \cosh \frac{E}{\kappa c^2} - \frac{1}{2\kappa^2 c^2}
  \, e^{\frac{E}{\kappa c^2} } \, \vec{p}^{\ 2}
 = 1 + \frac{M^2}{2\kappa^2}
 \end{equation}
One can consider the formula (\ref{2}) as describing the
deformation of classical energy-momentum dispersion relation
$\omega( \vec{p})= ( \vec{p}^{\ 2} + m_0^2 c^2 )^{1/2}$ for free
particles [3]
\begin{eqnarray}\label{3}
  \omega( \vec{p}) \to E_\kappa ( \vec{p}) & = & - \kappa c^2 \ln\left[1 +
  \frac{M^2}{2\kappa^2} - \sqrt{\left(1 +\frac{M^2}{2 \kappa c^2}\right)^2} +
  \frac{\vec{p}^{\ 2}}{\kappa^2 c^2}\right]
  \cr
&  = & - \kappa c^2\ln\left[\cosh(\frac{m_0}{\kappa}) - \sqrt{\sinh^2
(\frac{m_0}{\kappa}) +
  \frac{\vec{p}^{\ 2}}{\kappa^2 c^2}}\right]
\end{eqnarray}
where the value $M^2$ is related with the rest mass of particle as follows
[15]
\begin{equation}\label{4}
  M^2 = 2 \kappa^2 (\cosh \frac{m_0}{\kappa} -  1 ) \, .
\end{equation}
The quantum group structure of $\kappa$-deformed Poincar\'{e} algebra is
provided by nonprimitive  coproducts, which for threemomenta $P_i$ and energy
$E$ take the  following form
\begin{eqnarray}\label{5}
  \Delta P_i &= & P_i \otimes 1 + e^ { - \frac{E}{\kappa c^2} }
  \otimes P_i\, ,
  \cr
\Delta E &= & E \otimes 1 + 1 \otimes  E \, .
\end{eqnarray}
The $\kappa$-deformed relativistic framework is  described by dual
pair of Hopf algebras describing $\kappa$-deformed Poincar\'{e}
algebra and $\kappa$-deformed  Poincar\'{e} group [17-19].
Considering noncommutative translations $\widehat{x}_{\mu}$ of
$\kappa$-deformed Poincar\'{e} group as describing noncommutative
space-time coordinates one can show that
\begin{equation}\label{6}
\left[ \widehat{x}_0, \widehat{x}_i \right] = -\frac{i
\hbar}{\kappa c} \widehat{x}_i\, , \qquad \left[ \widehat{x}_i,
\widehat{x}_j \right] = 0 \, .
\end{equation}
The relations  (\ref{6}) describe $\kappa$-deformed Minkowski
space [17,9,3]. Further, by considering semidirect product of
$\kappa$-deformed Poincar\'{e} algebra and $\kappa$-Poincar\'{e}
group (so-called Heisenberg double [18])  one obtains
$\kappa$-deformed generalized phase space [18--19].

The aim of this note is  to consider the  possible definitions of
$\kappa$-deformed velocities and study their addition law. Due to
non-primitive coproduct (\ref{5}) one can introduce three
different  velocity formulae:

i) The one following from classical Hamilton equation
\begin{equation}\label{7}
\dot{X}_i = V_i = \frac{\partial E_\kappa ( \vec{p})}{\partial
p_i}\, .
\end{equation}
By considering  $\kappa$-deformed phase space in bicrossproduct
basis (see e.g. [18--20]) one confirms that the symplectic form
defining Hamiltonian equations (\ref{7}) are not deformed.

ii) Two other types of velocities are linked with  nonAbelian
addition law of three-momenta. By considering $P_i \otimes 1 =
p_i$, $1\otimes P_i = \delta^L p_i$ one can write
\begin{equation}\label{8}
\vec{p} + e^{-\frac{E}{\kappa c^2}}\delta^L \vec{p} \ = \ \vec{p}
+ \delta \vec{p}\Rightarrow \delta^L \vec{p} \ = \
e^\frac{E}{\kappa c^2} \delta \vec{p}\, .
\end{equation}
We define the left-covariant velocity as follows:
\begin{equation}\label{9} V_i^{L} =
\lim\limits_{\delta  p_i \to 0} \frac{E_\kappa(\vec{p}+
\delta\vec{p}) - E_\kappa(\vec{p})}{\delta^L p_i} =
e^{-\frac{E(\vec{p})}{\kappa c^2}}\frac{\partial E_\kappa (
\vec{p})}{\partial p_i} \ = \ e^{-\frac{E(\vec{p})}{\kappa c^2}}
V_i  \, .
\end{equation}
Using the assignment $P_i\otimes 1 = \delta^R p_i, 1\otimes P_i =
p_i$ one obtains
\begin{eqnarray}\label{10}\delta^R \vec{p} +
e^{-\frac{\delta E}{\kappa c^2}} \vec{p} \ = \ \vec{p} + \delta
\vec{p}\Longrightarrow  \delta^R \vec{p} \  = \ \delta \vec{p} +
\left(1- e^{-\frac{\delta E}{\kappa c^2}}\right) \vec{p} \cr = \
\left(1 + \frac{1}{\kappa c^2} \vec{p}\,\vec{\nabla}
E\right)\delta \vec{p}\, .
\end{eqnarray}
where assuming that momenta and velocities are parallel, i.e.
$\vec{p}\parallel \vec{V}=\vec{\nabla} E(\vec{p})$ we used the
relation
\begin{equation}\label{11}\vec{p} (\vec{\nabla}E\cdot
\delta \vec{p}_i) \ = \ (\vec{p}\cdot \vec{\nabla}E)\delta
\vec{p}_i\, .
\end{equation}
Using (10) one can introduce right-covariant velocity
\begin{equation}\label{12} V_i^{R} =
\lim\limits_{\delta  p_i \to 0} \frac{E_\kappa(\vec{p}+ \delta
\vec{p}_i) - E_\kappa(\vec{p})}{\delta^R p_i} \ = \ \left(1 +
\frac{1}{\kappa c^2}\vec{p}{\,} \vec{V}\right)^{-1}\frac{\partial
E_\kappa ( \vec{p})}{\partial p_i}\, .
\end{equation}
The velocities (\ref{7}) have been introduced in standard basis
still in 1992 by Bacry [21] (in bicrossproduct basis see [22]).
The velocities $V_i^{\ L}$ were introduced  in [23,24] and both
velocities (\ref{9}) and (\ref{12}) for massless case in [24] as
left and right group  velocities.

We shall show that the most interesting with its properties is the
velocity $V_i^R$ - it has classical velocity addition law, which
for parallel velocities $(0,0,V^R_1)$, $(0,0,V^R_2)$ looks as
follows
\begin{equation}\label{13}
V^{R}_{12}=  \frac{V^{R}_{1}+ V^{R}_{2}}{1 +
\frac{V^{R}_{1}V^{R}_{2}}{c^2}}\, .
\end{equation}
We shall discuss below all the three velocities described by
formulae (\ref{7}), (\ref{9}) and (\ref{12}) for arbitrary value
of mass parameter $M$ (see also (\ref{4})).

\section{Three Velocities - General Properties}

Three velocities (\ref{7}), (\ref{9}) and (\ref{12}) are related
with each other by the following formulae

\renewcommand{\theequation}{14\alph{equation}}
\setcounter{equation}{0}
\begin{eqnarray}\label{14a}
  V^{L}_{i} & = & e^{- \frac{E_{\kappa} (\vec{p} )}{\kappa c^3}} V_i\, ,
\\
\label{14b} V^R_i &= & \frac{V_i}{1+ \frac{1}{\kappa c^2} \vec{p}
\vec{\rm v}}\, ,
\end{eqnarray}
\renewcommand{\theequation}{\arabic{equation}}
\setcounter{equation}{14} where from (\ref{7}) and
(\ref{1}-\ref{2}) follows that\footnote{Further we denote $V=|
\vec{V}|$, $E=E_\kappa ( \vec{p})$ and $p^2 \equiv \vec{p}^{\ 2}
$.}
\begin{equation}\label{15}
  \vec{V} = \frac{\vec{p}}{ \frac{\kappa}{2} \left( 1 - e^{- \frac{2E}{\kappa c^2}} -
\frac{\vec{p}^{\ 2}}{\kappa^2 c^2} \right)} = \frac{\vec{p} \ e^\frac{E}{\kappa
c^2}}{\kappa\left[\cosh(\frac{m_0}{\kappa}) - e^{-\frac{E}{\kappa
c^2}}\right]}\, .
\end{equation}
One can calculate that$^{1}$
\begin{equation}\label{16}
V^2 = c^2 e^\frac{2E}{\kappa c^2}\left[1 -
\frac{\sinh^2(\frac{m_0}{\kappa})}{\left(\cosh (\frac{m_0}{\kappa}) -
e^{-\frac{E}{\kappa c^2}}\right)^2}\right]\,.
\end{equation}
and one obtains
\begin{equation}\label{17}
  \lim\limits_{E\to \infty} V  = \infty \, .
\end{equation}
In the interval  $M\leq E\leq \infty$ the function (\ref{16})
increases monotonically.

From the formulae (14) and (\ref{15}) one   gets
\begin{equation}\label{18}
  \vec{V}^L = \frac{e^{- \frac{E}{\kappa c^2}} \, \vec{p}}{ \frac{\kappa}{2}
\left( 1 - e^{-\frac{2E}{\kappa c^2}} - \frac{\vec{p}^2}{\kappa^2 c^2}\right)}
=  \frac{\vec{p} }{\kappa\left[\cosh(\frac{m_0}{\kappa}) - e^{-\frac{E}{\kappa
c^2}}\right]} \, .
\end{equation}
and after simple algebraic manipulation
\begin{equation}\label{19}
\vec{V}^R =  \frac{\vec{p}}{ \frac{\kappa}{2} \left( 1 - e^{-
\frac{2E}{\kappa c^2}} + \frac{\vec{p}^{\ 2}}{\kappa^2 c^2}
\right)} = \frac{\vec{p} \ e^\frac{E}{\kappa
c^2}}{\kappa\left[e^{\frac{E}{\kappa c^2}} -
\cosh(\frac{m_0}{\kappa}) \right]}\, .
\end{equation}
We get
\begin{eqnarray}\label{20}
(V^L )^2 & = & c^2 \left[1 - \frac{\sinh^2(\frac{m_0}{\kappa})}{\left(\cosh
(\frac{m_0}{\kappa}) - e^{-\frac{E}{\kappa c^2}}\right)^2}\right]  \, ,
\\
\label{21}
  (V^R )^2 & = &  c^2 \left[1 - \frac{\sinh^2(\frac{m_0}{\kappa})}{\left(\cosh
(\frac{m_0}{\kappa}) - e^{\frac{E}{\kappa c^2}}\right)^2}\right]\, .
\end{eqnarray}

One can show:

i) for all energies $(V^L)^2\leq c^2$ and $(V^R)^2\leq c^2$

ii) for $M<E< \infty$ we get $\frac{d(V^L)^2}{dE}
> 0$ and $\frac{d(V^R)^2}{dE}  > 0$, i.e. both functions
(\ref{20}) and (\ref{21}) are monotonic.

iii) if $M=0$ (equivalent to $m_0 = 0$) both velocities $V^L_i$
and $V^R_i$ have classical absolute value  i.e.  $(V^L)^2
=(V^R)^2=c^2$ \,\footnote{The observation iii) has been made also
in [24].}

\section{Right Group Velocity - Addition Formula}

Let us write the formula (\ref{19}) as follows:
\begin{equation}\label{22}
  \vec{p}= \kappa \, B ( m_0 , E)
  \vec{V}^R\, ,
\end{equation}
where
\begin{equation}\label{23}
  B( m_0 , E) = 1 - \cosh(\frac{m_0}{\kappa}) \ e^{-\frac{E}{\kappa c^2}}\, .
\end{equation}
The function $B( m_0 , E)$ enters into the $\kappa$-deformed Lorentz
transformations determined by the boost parameter $ \vec{\alpha} = \alpha
\vec{n}$ ($\vec{n}^{\ 2} = 1$) [25,16]

\renewcommand{\theequation}{24.\alph{equation}}
\setcounter{equation}{0}
\begin{eqnarray}\label{24a}
E( \alpha )& = & E + \kappa c^2 \ln W(E, \vec{n} \vec{p}{\ ;}
\alpha )\, ,
\\
\label{24b} \vec{p}(\alpha)& = & W^{-1} (E, \vec{n} \vec{p}{\
;}\alpha) \left\{ \vec{p} + \left[ (\vec{n}\vec{p})(\cosh \alpha -
1 )\right.\right. \cr && \left. \left. - \kappa \, c B(m_0 , E)
\sinh \alpha\right]\vec{n}\right\}\, .
\end{eqnarray}

\renewcommand{\theequation}{\arabic{equation}}
\setcounter{equation}{24} where
\begin{equation}\label{25}
  W(E, \vec{n}\vec{p}{\ ;} \alpha)
  = 1 - \frac{1}{\kappa c} (\vec{n}\vec{p})\sinh \alpha +
  B( m_0 , E) ( \cosh \alpha - 1)\, .
\end{equation}
The function (\ref{25}) satisfies the relation
\begin{equation}\label{26}
  W(E, \vec{n}\vec{p}{\ ;}\alpha)
  = \frac{B(m_0 , E)- 1}{B(m_0 , E(\alpha))-1}\, .
\end{equation}
Let us define the velocity $ \vec{W}^R$ as the velocity (\ref{22})
in the frame Lorentz-transformed by the boost parameter $
\vec{\alpha}= \alpha \vec{n}$.

Using (\ref{22}) we have the formula
\begin{equation}\label{27}
\vec{p}(\alpha) = \kappa B( m_0 ,E(\alpha)) \overrightarrow{W}^R\,
.
\end{equation}
From (\ref{27}) and (\ref{24b}) one gets
\begin{eqnarray}\label{28}
\overrightarrow{W}^R& =& \frac{B( m_0 , E)}{W(E , \vec{n}\vec{p}{\ ;}\alpha)+
B( m_0 , E) -1} \cr\cr &&\cdot \left\{ \vec{V}^R + \vec{n}\left[
(\vec{n}\vec{V}^R) (\cosh \alpha -1) - c \sinh \alpha \right]\right\}\, ,
\end{eqnarray}
and further we obtain
\begin{equation}\label{29}
  \overrightarrow{W}^R =
  \frac{\vec{V}^R + \vec{n}[(\vec{n}\vec{u})(\cosh \alpha -1) - c\sinh \alpha]}
  {\cosh \alpha [1 - \frac{1}{c}(\vec{n} \vec{u})
\tanh \alpha]}\, .
\end{equation}
Introducing the relative velocity $\vec{u}$ of two $\kappa$-deformed Lorentz
frames, one at rest ($\alpha = 0$) and second described by boost parameter
$\vec{\alpha} = \vec{n} \alpha$
\begin{equation}\label{30}
  \vec{u} = - c\vec{n} \tanh \alpha \, ,
\end{equation}
the relation (\ref{29}) can be written in the form (see e.g. [26])
\begin{eqnarray}\label{31}
  \overrightarrow{W}^R &= & \left( 1 +  \frac{\vec{V}^R
  \vec{u}}{c^2}\right)^{-1}
  \left\{ \vec{V}^R \left( 1 - \frac{u^2}{c^2}
  \right)^{1/2} \right.
  \cr
  &&
  \left.
  + \,  \vec{u} \left[  1 + \frac{\vec{V}^R
  \vec{u}}{u^2}  - \frac{\vec{V}^R
  \vec{u}}{u^2} \left( 1 - \frac{u^2}{c^2} \right)^{1/2}\right]
  \right\}\, .
\end{eqnarray}
The formula (\ref{31}) describes the general Einsteinian
composition law of two arbitrary three-velocities $\vec{V}^R,
\vec{u} $, which for parallel velocities $\vec{V}^R \| \vec{u} $
reduces to the formula (\ref{13}).

It should be stressed that basic ingredient in the derivation of
classical addition law (\ref{31}) is the relation (\ref{22}). If
we use other two formulae (\ref{15}) or (\ref{18}) for deformed
velocities, analogous reasoning leads to the deformation of
classical addition formulae (\ref{13}) and (\ref{31}).

To complete the argument we shall show that the definition
(\ref{30}) is equivalent to the relation (\ref{22}). Indeed, let
us solve the relation $\vec{p}(\alpha)=0 $ describing the
transformation from the moving system with nonvanishing momentum
$\vec{p}(\alpha=0) $ to the rest system with $\vec{p}=0 $ ($\alpha
\neq 0$). From (\ref{24b}) one gets
\begin{equation}\label{32}
\vec{p}(\alpha)=0 \Longrightarrow \vec{p} + ( \vec{n}\vec{p})(\cosh \alpha -1 )
- \kappa c \, B(m_0 , E) \sinh \alpha = 0 \, .
\end{equation}
Further if $ \vec{n}\| \vec{p}$ one obtains $(p  \equiv | \vec{p}|)$
\begin{equation}\label{33}
  p \cosh \alpha - \kappa c \, B(m_0 , E) \sinh \alpha = 0 \, .
\end{equation}
We see that inserting in (\ref{33}) the velocity $u$ from formula
(\ref{30}) ($u = c\tanh \alpha $) we obtain the relation
(\ref{22}) ($p = \kappa B(m_0 , E)u$).

\section{Velocity and Noncommutative Space-Time}

In order to describe velocity formula  in noncommutative
space-time one should use the corresponding deformed Hamiltonian
formalism. For $\kappa$-de\-formed relativistic phase space such a
framework has been proposed in [3]\footnote{see formulae
  (1.5--1.10) and  4.2 in [3].}

The $\kappa$-deformed noncommutative phase space kinematics is
determined by basic Poisson brackets of relativistic phase space
variables $Y_A = (x_\mu, p_\mu)$ given in bicrossproduct basis by
the following relations [18,19]:
\begin{eqnarray}\label{34}
\left\{  p_i, x_j \right\} & = & \delta_{ij}\, , \cr \left\{ p_0,   x_0
\right\} & = & - 1\, , \cr \left\{ p_0, x_i \right\} & = &
 0\, ,
\cr \left\{ x_0, x_i \right\} & = & \frac{x_i}{\kappa c}\, , \cr
\left\{ x_0, p_i \right\} & = & - \frac{p_i}{\kappa c}\, , \cr
\left\{ p_\mu , p_\nu \right\} & = & 0 \, .
\end{eqnarray}
Writing down (\ref{34}) in compact form
\begin{eqnarray*}\label{nolu*}
  \left\{  Y_A , Y_B\right\} = \omega^{(\kappa)}_{AB} (x,p) \, .
\end{eqnarray*}
we obtain the $\kappa$-deformed Hamilton equations describing evolution with
respect to the parameter $s$
\begin{equation}\label{35}
\frac{dY_A}{ds} = \omega^{(\kappa)}_{AB} \frac{\partial {\cal
H}^{(\kappa)}}{\partial Y_B}\, .
\end{equation}
where $ {\cal H}^{(\kappa)}= {\cal H}^{(\kappa)}(x,p)$ determines
the dynamics. Assuming translational invariance one should take
one-particle Hamiltonian as ${\cal H}^{(\kappa)} = {\cal
H}^{(\kappa)}(p_0 , \vec{p})$ and explicitely from (\ref{35}) one
gets
\renewcommand{\theequation}{36\alph{equation}}
\setcounter{equation}{0}
\begin{eqnarray}\label{36a}
\frac{dx_i}{ds}& = & - \frac{\partial {\cal H}^{(\kappa)}}{\partial p_i}\, ,
\\
\label{36b}  \frac{dx_0}{ds} & = & -\frac{1}{\kappa c} \, p_i
\frac{\partial {\cal H}^{(\kappa)}}{\partial p_i} + \frac{\partial
{\cal H}^{(\kappa)}}{\partial p_0}\, .
\end{eqnarray}
and $ \frac{dp_\mu}{ds} = 0$ i.e. $p_{\mu}$ are $s$-independent.

\renewcommand{\theequation}{\arabic{equation}}
\setcounter{equation}{36}

The physical interpretation of the parameter $s$ depends on the choice of the
Hamiltonian $ {\cal H}^{(\kappa)}$. We can consider the following two basic
cases:

i) The Hamiltonian $ { H}^{\kappa}$ describes the energy
dispersion relation $ E=E_{\kappa} (\vec{p} )$ by means of the
formula $ {\cal H}^{(\kappa)} \equiv { H}^{\kappa}(p_0 , \vec{p}
)= c p_0 - E_\kappa (\vec{p} )$. In such a case if $\kappa \to
\infty$ (standard relativistic framework) i.e. $H^{\infty}(p_0 ,
\vec{p} )= c p_0 - c(\vec{p}^{ \ 2} + m^2_0 c^2 )^{1/2} $ one can
identify the parameter $s$ with time variable ($\frac{dx_0}{ds}= 1
\longleftrightarrow x_0 = s+ const. $) and one obtains the
standard velocity formula
\begin{equation}\label{37}
\frac{dx_i}{dx_0} \ = \ \frac{p_i}{(\vec{p}^{\ 2} + m^2_0
c^2)^{1/2}} \, .
\end{equation}
If $\kappa \neq 0$ such identification is not possible, because
from (\ref{36b}) it follows that
\begin{equation}\label{38}
\frac{dx_0}{ds}= c - \frac{1}{\kappa c} \, p_i \frac{\partial
H^\kappa}{\partial p_i}\, .
\end{equation}
and further we get
\begin{equation}\label{39}
\frac{dx_i}{dx_0} \ = \ \frac{\frac{dx_i}{ds}}{\frac{dx_0}{ds}} \
= \ \frac{- \frac{\partial H^{\kappa}}{\partial p_i}}{c\left(1 -
\frac{1}{\kappa c^2} p _i \frac{\partial H^{\kappa}}{\partial
p_i}\right) }\, .
\end{equation}
i.e after using  (\ref{7}) we obtain our favoured velocity formula
(\ref{14b}).

ii) One can use also as the Hamiltonian ${\cal H}^{(\kappa)} $ the
$\kappa$-invariant mass Casimir  by assuming that $ {\cal
H}^{(\kappa)}\equiv M^2 c^4$. In classical relativistic case $
{\cal H}^{(\infty)}= E^2  - c^2 \vec{p}^{\ 2}= c^2\left(p_0^2 -
\vec{p}^{\ 2}\right) $ the parameter $s$
 corresponds to Poincar\'{e}-invariant length parameter. One gets
\begin{equation}\label{40}
\frac{dx_i}{ds}= 2 c^2 p_i \, , \qquad \qquad \frac{dx_0}{ds}= 2
c^2 p_0 \, .
\end{equation}
and on the mass-shell ( ${\cal H}^{(\infty)} = m^2_0 c^4 = const.
$) we obtain the formula (\ref{37}). In general case ($\kappa \leq
\infty $) the deformed mass shell condition (2) can be written as
identity ${\cal H}^{(\kappa)}(E_\kappa(\vec{p}), \vec{p})\equiv
M^2 c^4$ where $E_\kappa(\vec{p})$ is given by (3). One obtains
\begin{equation}\label{41}
\frac{\partial {\cal H}^{(\kappa)}}{\partial p_0} \,
\frac{\partial { E}_{\kappa}}{\partial p_i} + \frac{\partial {\cal
H}^{(\kappa)}}{\partial p_i} = 0 \Rightarrow \frac{\partial {\cal
H }^{(\kappa)}}{\partial p_i} \ = \ - V_i\, \frac{\partial {\cal
H}^{(\kappa)}}{\partial p_0} \, .
\end{equation}
After inserting (\ref{41}) into (\ref{36a}-\ref{36b}) one gets
\begin{equation}\label{42}
\frac{dx_i}{ds} \ = \ V_i\,\frac{\partial {\cal
H}^{(\kappa)}}{\partial p_0}\, , \frac{dx_0}{ds} \ = \ \left(1 +
\frac{1}{\kappa c} p_i V_i\right)\frac{\partial {\cal
H}^{(\kappa)}}{\partial p_0} \, .
\end{equation}
and we obtain again the formula (\ref{14b}).

\section{Final Remarks}
Following the philosophy advocated firstly by Majid [27]
quantum-deformed space-time kinematics describes generalized
symmetries and noncommutative geometries at ultra-short Planck
scales. The real challenge is to find observable physical effects
caused by such modification of short distance behaviour of quantum
phenomena. The study presented here has much more modest aim: it
provides a contribution to the description of kinematics obtained
in the framework of deformed quantum theories.

It should be pointed out that such a kinematical framework
requires still several problems to be solved, in particular
understanding the relation between $\kappa$-deformed kinematics
and description of macroscopic bodies. The basic question is to
understand how the $\kappa$-effects cancel if we consider very
large summ of elementary objects described by deformed
$\kappa$-kinematics. We would like to stress that in such a case
we should obtain for macroscopic bodies the classical relativistic
kinematics.

 \end{document}